\documentclass[a4paper,12pt]{article}
\usepackage{theorem,amssymb}
\usepackage{graphicx}
\def\Tr{\mathrm{Tr}}

\newtheorem{tetel}{Theorem}

\newtheorem{pelda}{Example}

\newcommand{\<}{\langle}
\renewcommand{\>}{\rangle}

\def\bbbr{{\mathbb R}}
\def\bbbc{{\mathbb C}}

\def\Tr{\mathrm{Tr}\,}

\def\Det{\mbox{det}\,}

\def\iH{{\cal H}}

\def\ep{\varepsilon}

\begin{document}
%\rightline{\today }
%\vskip 1cm
\centerline{\large {\bf Optimal quantum state tomography with known parameters}}
%\bigskip
%\centerline{\LARGE {\bf }}
\bigskip
\centerline{D\'enes Petz and L\'aszl\'o Ruppert}
\bigskip
\centerline{Department of Mathematical Analysis}
\centerline{Budapest University of Technology and Economics} 
\centerline{Egry J\'ozsef u.~1., Budapest, 1111 Hungary}

\begin{abstract}

It is a well-known fact that the optimal POVM for quantum state tomography is the symmetric, informationally complete, positive operator valued measure (SIC-POVM). We investigate the same problem only in the case when there are some a priori information about the state, specifically when some parameters are known. In this paper we mainly focus on solving a 3-dimensional optimization problem, which gives us a non-trivial example for the so-called conditional SIC-POVMs, a straightforward generalization of the concept of SIC-POVMs. We also present other special cases to show further applications of the proposed numerical methods and to illustrate the complexity of this topic.

\end{abstract}

\section{Introduction}

State estimation is a fundamental problem in the field of quantum
information theory and it can be considered as one of the
foundational issues in quantum mechanics \cite{Nielsen-book,PDbook}.
The problem may be traced back to the seventies \cite{Helstrom},
the interest in a thorough mathematical analysis of the
quantum tomography procedures has been flourishing recently
\cite{tomography,Nunn2010,Scott06,Teo}.

In statistics the accuracy of the estimation can be quantified by
the covariance matrix. The matrices are typically not
comparable by the positive semi-definiteness, hence if different
estimation schemes are compared, the determinant of the covariance matrix can be used instead. 
This approach can be found in references \cite{PeHa:2007, PeHa:2008}, their main result was
that the complementary von Neumann measurements are optimal. When all parameters of the
density matrix are obtained from a single measurement, then a
symmetric informationally complete POVM turns out to be optimal \cite{renes}.
A similar result was obtained earlier by Wootters and
Fields \cite{WoFi} for von Neumann measurements and by Scott \cite{Scott06} for POVMs,
but optimality had a different formulation in both cases.

Some a priori information about the state can be given in various ways, the most popular subject in this field is state discrimination: when we know that the system is in one of several given states, and we should figure out which one it is \cite{dar05}. Beside knowing the possible states we can have an a priori probability distribution on the true state, too. This idea was used in \cite{demko} to obtain the optimal phase estimation. There the given states do not construct a discrete set, instead, they are searching among all pure states. In our setup we use a similar assumption, we know that the state is in a given subset of the whole state space (some parameters of the state are given) and optimize a quantity introduced in \cite{PR}. 

The problem is examined thoroughly through a 3-dimensional example, but we give results for other examples, too. 
The optimization is hard to handle analytically, therefore numerical methods are proposed to find the optimal POVM.
We achieved quite fast convergence and we can even give analytic solutions of the state estimation problems. The results suggest a strong pattern for optimal POVMs, which leads us to the generalization of the concept of SIC-POVMs.

The rest of the paper is organized in the following way: In Section 2 we give a short overview of the used concepts. Then, we define the efficiency of our estimation that we want minimize (Section 3). In Section 4 we give an algorithm which solves the minimization problem numerically. Section 5 gives the results for various examples. 
Finally, we discuss the results, introduce the definition of conditional SIC-POVMs and draw the conclusions (Section 6).

%%%%%%%%%%

\section{Basic concepts}\label{optprob}

In this section we will give an overview of the concepts used in this article, for a more detailed description see \cite{Nielsen-book,PDbook}.

\subsection{Quantum states }

An $n$-dimensional quantum state can be described by using density matrix $\rho \in M_n (\bbbc)$ fulfilling the following conditions:
$$
\rho \ge 0 \quad  \textrm{and} \quad \Tr \rho=1.
$$

Let us denote the n-dimensional generalized Pauli matrices with $\sigma_i, \quad(i\in \{0, 1, \dots, n^2-1\})$, where $\sigma_i\ge 0$, $\Tr (\sigma_i \sigma_j)=\delta_{i,j}$, that is, we have an orthonormal basis on the positive matrices and use the abbreviations $\sigma=\{\sigma_i: 1 \le i \le n^2-1\}$ and $\sigma_0=I/\sqrt{n}$.

Then the density matrix $\rho \in M_n (\bbbc)$ will have $n^2-1$ real parameters, namely the elements of $\theta \in \bbbr^{n^2-1}$, which can be referred to as the generalized Bloch-vector:
$$
\rho=\frac{I}{n}+\theta \cdot \sigma=\frac{I}{n}+\sum_{i=1}^{n^2-1} \theta_i \sigma_i.
$$

In our paper we will use the assumption that there are exactly $N$ unknown parameters (we can suppose that the unknown parameters are: $\theta_1, \theta_2, \dots, \theta_N$) and our aim is to get the most efficient estimation of these parameters. 

\subsection{Quantum measurements}

For measurements, we will use a single finite POVM: $E=\{E_1,E_2,\dots E_m\}$, which satisfies the conditions 
$$\sum_{j=1}^m E_j=I \quad \textrm{and} \quad E_j\ge 0.$$

The first equality shows that the $E_j$-s are not completely independent, so if we want to have $N$ independent POVM elements to estimate the $N$ parameters of the state, the POVM $E$ has to have at least $m=N+1$ elements. 

Then we use a similar parametrization as in the case of the quantum states:
$$
E_1=a_0^{(1)} (I+ a^{(1)} \cdot \sigma), \quad  E_2=a_0^{(2)} (I+ a^{(2)} \cdot \sigma), \quad \dots \quad
E_{N+1}=a_0^{(N+1)} (I+ a^{(N+1)} \cdot\sigma)\,,
$$
where $a_0^{(j)} \in \bbbr$,  $a^{(j)} \in \bbbr^{n^2-1}$.

The positivity conditions for $E_j$ are: $a_0^{(j)}\ge 0$ and $a^{(j)} \in \mathcal{P}$, with 
$\mathcal{P}$ denoting the set of $a$ vectors which satisfy $I+ a \cdot \sigma>0$.

Then the probability of getting an outcome related to $E_j$ is
$$
p_j=\Tr( E_j \rho)=a_0^{(j)} + a_0^{(j)} \< a^{(j)}, \theta\> \quad (j=1,2,\dots N).
$$
In matrix notation we have
\begin{equation}\label{E:sol1}
\left[\matrix{p_1 \cr \vdots \cr p_N }\right]=
\left[\matrix{a_0^{(1)} \cr \vdots \cr a_0^{(N)} }\right]+T
\left[\matrix{ \theta_1 \cr \vdots  \cr \theta_N }\right]
\end{equation}
with the matrix
$$
T:=\left[ \begin{array}{ccc} a_0^{(1)}a_1^{(1)}& \dots & a_0^{(1)}a_N^{(1)}  \\
\vdots &  & \vdots  \\
a_0^{(N)} a_1^{(N)}& \dots & a_0^{(N)} a_N^{(N)} \end{array} \right].
$$

\subsection{Quantum state estimation and its efficiency}\label{eff}

Let us assume that we have many identical copies of $\rho$ and repeat the previously described measurement on each of them.
If $\nu_1,\dots,\nu_N$ are the relative frequencies of the outcomes related to $E_1,~E_2, \dots$ and $E_N$ respectively, then from (\ref{E:sol1}) we can obtain the state estimate 
\begin{equation}\label{E:sol2}
\left[\matrix{ \hat\theta_1 \cr \vdots  \cr \hat\theta_N }\right]=T^{-1}
\left[\matrix{\nu_1-a_0^{(1)} \cr \vdots \cr \nu_N-a_0^{(N)}}\right].
\end{equation}
It is easy to see that this is an unbiased and efficient estimator gained from the measurement outcomes on the unknown parameters of the quantum state.

The covariance matrix of the random variable $\hat\theta$ is
$$
V(\rho)=\left[\mathbb{E}(\hat\theta_i-\theta_i)(\hat\theta_j-\theta_j)\right]_{i,j\in\{1, \dots, N\}}=T^{-1}\,W\,(T^{-1})^*
$$
where $W$ is the covariance matrix of the random variable $(\nu_1,\dots,\nu_N)$.  If $M$ is the number of measurements and $M_j$ is the number of outcomes related to $E_j$, then $(\nu_1,\dots,\nu_N)=\frac{1}{M}\cdot (M_1, M_2, \dots M_N)$, and since $(M_1, M_2, \dots M_N)$ have a multinomial distribution we have
$$
W=\frac{1}{r}\left[\matrix{p_1(1-p_1) & - p_1 p_2 &\dots &  - p_1 p_N\cr
 - p_1 p_2 & p_2(1-p_2) &\dots&  - p_2 p_N\cr
\vdots&\vdots &  &\vdots\cr
 - p_1 p_N &  - p_2 p_N &\dots & p_N(1-p_N)}\right].
$$

\subsection{Complementarity and symmetric measurements}

The heuristic concept of complementarity was born together with quantum theory.
A mathematical definition is due to Accardi \cite{Acc} and Kraus \cite{Kr}. Let
$\iH$ be an $n$-dimensional Hilbert space. Let the observables $A$ and $B$ have
eigenvectors $e_1,e_2,\dots, e_n$ and $f_1,f_2,\dots, f_n$ that are orthonormal
bases. Then $A$ and $B$ are complementary if
\begin{equation}\label{E:c}
|\< e_i, f_j\>|^2=\frac{1}{n}\qquad (1 \le i,j \le n).
\end{equation}
If this condition holds then the two bases are also called mutually unbiased.

Complementarity can be generalized to the case of POVMs. The POVMs $\{E_1,E_2,\dots, E_k\}$
and $\{F_1,F_2,$ $\dots, F_m\}$ are complementary if
\begin{equation}\label{E:c2}
\Tr E_i F_j=\frac{1}{n}\Tr E_i \, \Tr F_j \qquad (1 \le i \le k, \quad 1 \le j \le m).
\end{equation}
This is equivalent to the orthogonality of the traceless parts:
$$
E_i - \frac{\Tr E_i}{n} I \quad \perp \quad F_j - \frac{\Tr F_j}{n} I,
$$
so we will use the expression quasi-orthogonal when we use this property of complementary operators.
%The concept can be extended to subalgebras $\iA_1,\iA_2 \subset \Mn$. $\iA_1$ and $\iA_2$
%are complementary if  $A_1 \in \iA_1$ and $A_2 \in \iA_2$ are quasi-orthogonal (that is, the traceless part of matrices are orthogonal).
An overview about complementarity can be found in \cite{jmp}.

The symmetric informationally complete POVM is a popular subject in quantum tomography \cite{sic3dim,renes,zauner}. 
A SIC POVM $\{E_i\,:\, 1 \le i \le k\}$ of an $n$-level system is described by a  
set of projections $P_i=|h_i\>\<h_i|$ ($1 \le i\le k$) such that
\begin{equation}\label{SIC}
\sum_{i=1}^k P_i=\lambda I \quad \mbox{and} \quad \Tr P_i P_j=\mu \quad (i\ne j),
\end{equation}
where $k=n^2$, $\lambda=n$ and $\mu=1/(n+1)$. The existence
of a symmetric informationally complete POVM (SIC POVM) is not known for a general dimension
$n$ \cite{scott}.

\section{The optimization problem}

To obtain the optimal measurement setup we need to quantify the error of our estimation, and we want to minimize it over the all possible POVMs. In Section \ref{eff} we defined an estimator and we calculated its covariance matrix, but the problem is that matrices are in general not comparable, so we need a function from $f:\mathbb{R}^{N\times N} \rightarrow \mathbb{R}$. A common technique is  to take the integral with respect to the Haar measure on the unitaries (e.g., \cite{Scott06}), this way we can in some sense symmetrize the result and handle the a priori knowledge about the state. 
From \cite{PR} we can conclude that the best choice of $f$ is to take the determinant of the covariant matrix, and the correct order is to take the average first and the determinant after that.

Let $H$ be the set of all possible states with the known parameters of the quantum state.
To get the average mean quadratic error matrix we integrate $V$ with respect to the Haar-measure on the unitaries  ($\mu$), but restricted to $H$, this way we only average the states with the given known parameters. So the average of the covariance matrix is
$$
\int \chi_{H}(U\rho U^* ) V(U \rho U^*)\, d\mu(U)=T^{-1}W_0(T^t)^{-1},
$$
where $\chi$ is the characteristic function, and
\begin{equation}\label{int}
W_0=\int \chi_{H}(U\rho U^* ) W(U\rho U^* )\, d\mu(U)
\end{equation}

The next step to take the determinant:
\begin{equation}\label{aim}
DACM=\Det\left(T^{-1}W_0(T^t)^{-1}\right)=\frac{\Det(W_0)}{\Det^2(T)}
\end{equation}
and we want to minimize the determinant of the average covariance matrix ($DACM$) for all the possible POVMs to get the optimal measurement.

The essence of \cite{PR} is that this quantity can be applied successfully for different optimization problems. For instance it is shown for the n-dimensional case that if there are no known parameters, i.e., we want a full state estimation, then the optimal POVM is the SIC-POVM. This is not a surprising result, however, the following Theorem is non-trivial and contains some interesting questions for the general case:

\begin{tetel}\label{T1}
In the qubit case the optimal POVM for the unknown $\theta_1$ and $\theta_2$ can be
described by the projections $P_i$ $(i=1,2,3)$:
$$
E_i=\frac{2}{3}P_i, \quad \sum_{i=1}^3 P_i=\frac{3}{2}I,
\quad \Tr \sigma_3 P_i=0, \quad \Tr P_iP_j= \frac{1}{4}
\quad \mbox{for}\quad  i \ne j.
$$
\end{tetel}

So this result is in some sense the combination of symmetricalness and complementarity. Both this result and the SIC-POVM case are proven analytically, but interestingly, the latter was more difficult to prove, since the total symmetricalness made it easier to prove even in the general, $n$-dimensional case. But the simplicity of $M_2(\bbbc)$ made the calculations feasible, so the question is, whether there is a similar object in higher dimensions or not. Actually, even in the 3-dimensional case, the precise mathematical argument is rather complicated and so the present approach is based on numerical methods.

\section{The algorithm}\label{alg}

In this section, we will show a method for solving the previously described optimization problem numerically using an example setting.
Specifically, let us assume that we have a 3-dimensional system, i.e., a qutrit ($n=3$), and we know the diagonal entries of the density matrix $\rho$.

%For practical reasons, we assume, that $a_0^{(j)}=1/7$ for all $j$. This is not a necessary condition, but the simulations converge much faster if we do so.

The first problem that emerges when implementing this minimization problem is the calculation of the objective function (\ref{aim}) for a given POVM. This is difficult because formula (\ref{int}) is quite abstract, but we can give a good approximation. We parametrize $M_3(\mathbb{C})$ using the Gell-Mann matrices, we use a dense enough grid on the parameter space $\bbbr^8$ and check for each grid-point whether it is an element of $H$. Actually, the Bloch-vector has only 6 parameters, since the diagonal entries of the states are known. The actual calculation consists simply of checking for all grid points the positive definiteness of the matrix determined by the actual generalized Bloch-vector. Then we cluster the grid points of $H$ according to their eigenvalues: we partition the interval $[0,1]$, and two states will belong to the same cluster if their eigenvalues are in the same cells. We choose one cluster, this means all the states with the ``same'' eigenvalues (i.e., achievable states using unitary transformations) and we take the sum of $W$ in these points. Let us note that we do not use a normalized measure either here, or in (\ref{int}), since it is not necessary: we get the same optimization problem up to a constant factor. Another remark is that if we choose a small cluster, the computation will be less precise than for a large one, but much faster. 

The second problem is how to select new POVMs to get better and better estimations. We choose an arbitrary initial point in the interior of the state space and in each step, we take a new random POVM by perturbing the parameters using a normal distribution with a given variance.  This means that for each $E_j, ~(j=1,...6)$, we calculate $\hat a^{(j)}=a^{(j)}+\mathcal{N}_8(0,s(t))$, we repeat the random realization of normal vectors for $E_1$, while $\hat  a^{(1)}$ will determine a positive matrix, then we continue the realization with $E_2$, and so on. The variance of normal distribution ($s(t)$) is decreasing in time, first we need a larger variance for faster convergence, but near the boundary of the state space of POVMs we will easily get negative eigenvalues if the disturbance is too high. If we have a new Bloch vector for all the 6 POVM elements, we take all the variation of $a^{(j)}$ and $\hat a^{(j)}$ ($\tilde a^{(j)} \in \{a^{(j)}, \hat a^{(j)}\}$), and we check for all the $2^6=64$ cases whether the correlated $E_7=I-E_1-E_2-\dots-E_6$ will be a physically possible state or not. Then we go through the valid POVMs and we use simulated annealing  \cite{Metropolis} for this series of POVMs. Let the current best POVM be $E$ and the next in the line to check is $\tilde E$, then we change the best POVM to $\tilde E$ with probability:
$$
\mathbb{P}(E \rightarrow \tilde E)=\frac{1}{1+\exp \left( \frac{log( DACM(\tilde E))-log (DACM (E))}{T} \right)},
$$
 where $T$ is the so-called temperature. For high temperatures, the probabilities are close to $1/2$ so the optimal POVM can roam freely, but for low temperatures, we change the current best POVM only if the new POVM is really better. This transition probability determines a special kind of Glauber dynamics, so there is a good chance that it will converge to the global optimum. The reason why we use simulated annealing instead of simply selecting the best POVM from the line is, because otherwise the algorithm tends to set in one direction and it only converges to the boundary of the state space. The simulated annealing is useful here because it can change this path by overcoming potential barriers. Also we increase the temperature from time to time to help escape from local optima.

\section{Results}\label{results}

\subsection{The 3-dimensional example}\label{3dim}

\begin{figure}[!ht]
\begin{center}
\begin{footnotesize}
\begin{tabular}{cc}
 \input{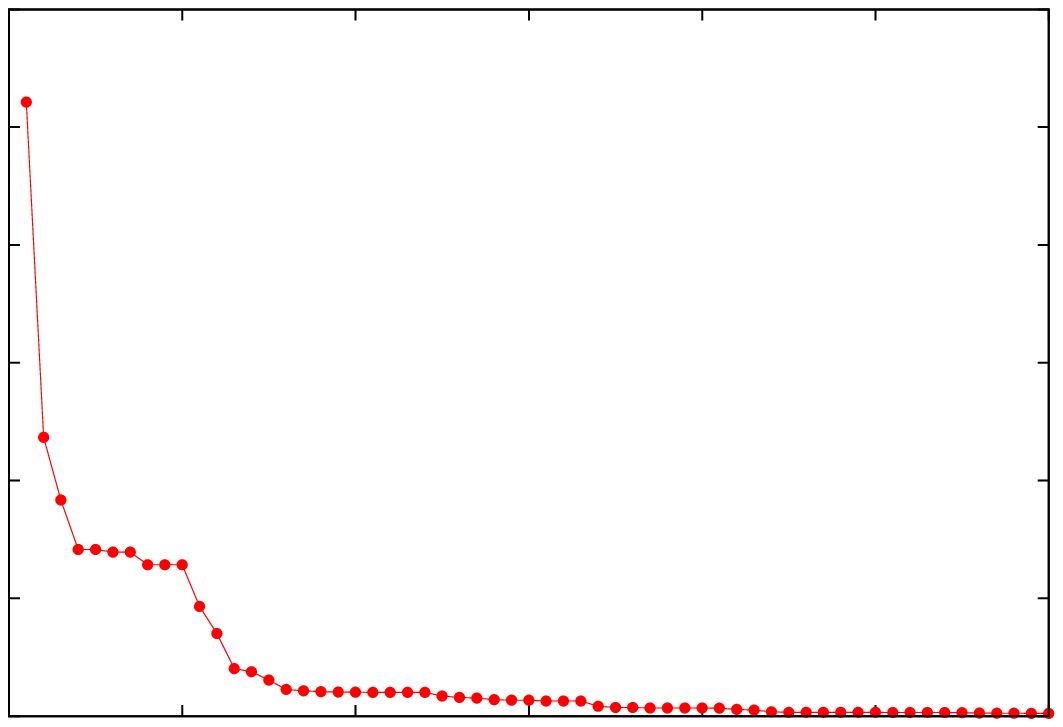} & \input{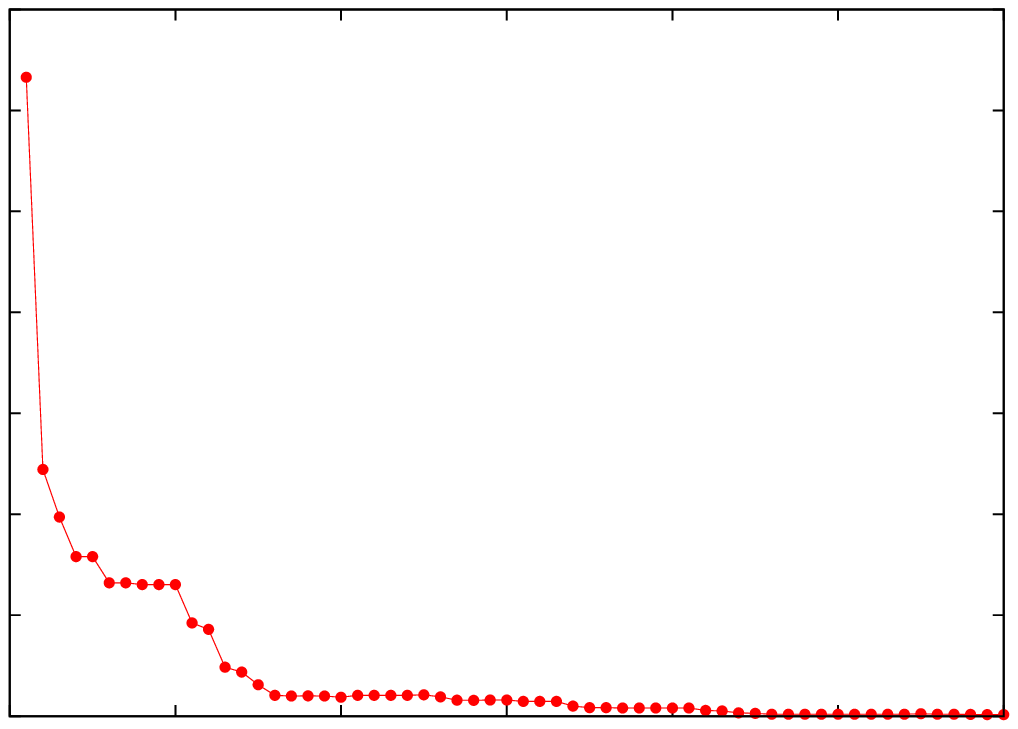}
\end{tabular}
\end{footnotesize}
\caption{The logarithm of $DACM$ and $\sigma$ as the function of the number of steps}
\label{fig:szor}
\end{center}
\end{figure}
      
\begin{figure}[!ht]
\begin{center}
\begin{footnotesize}
\begin{tabular}{cc}
 \input{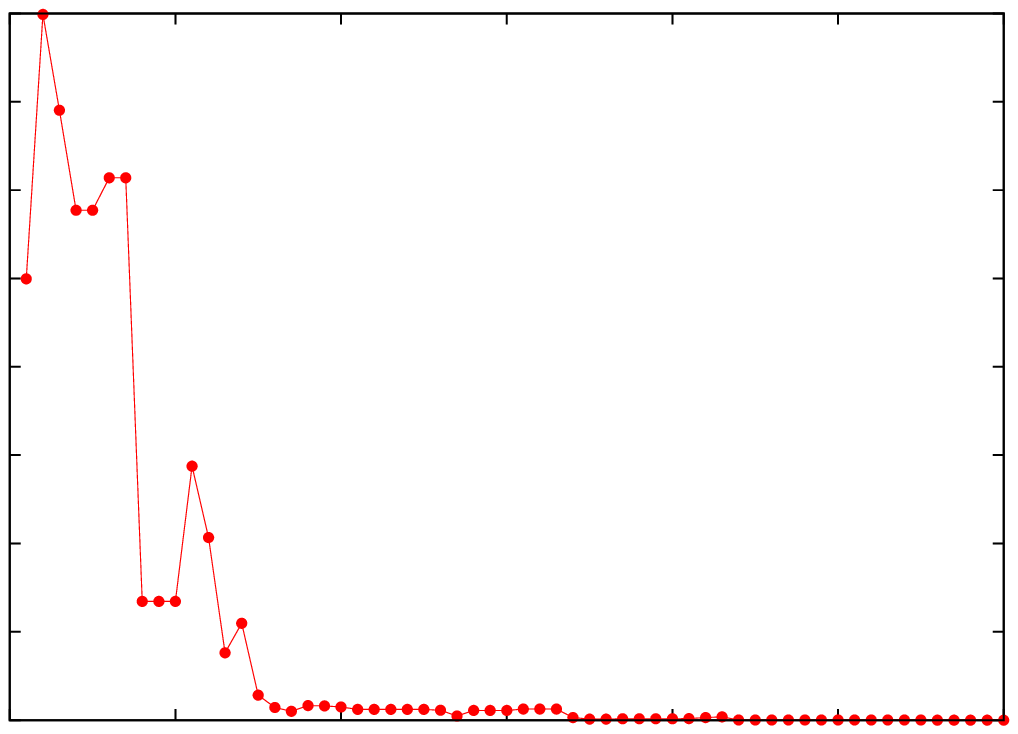} & \input{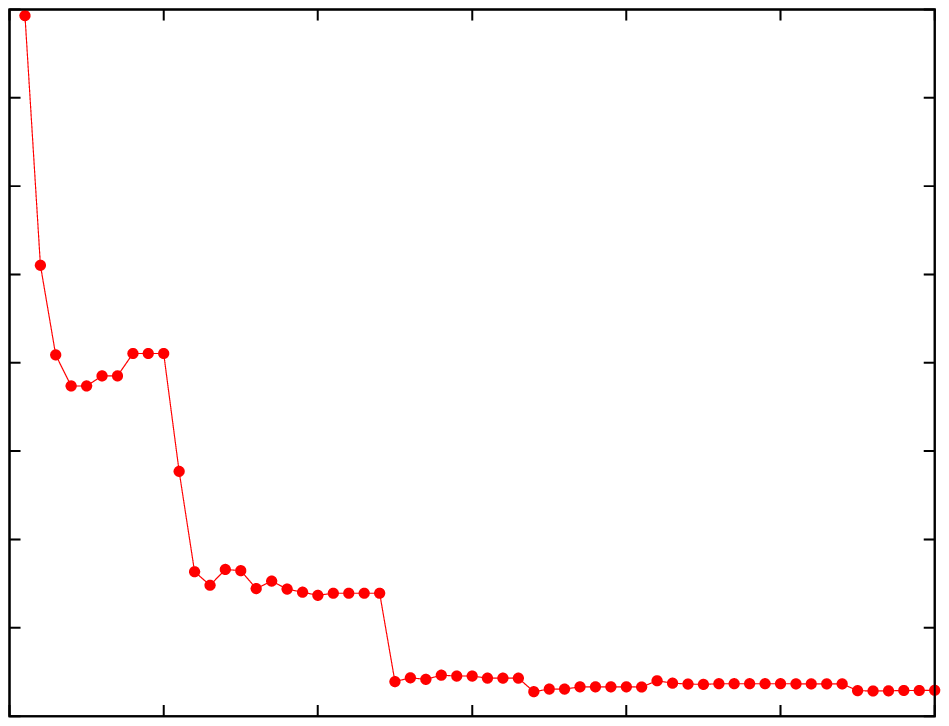}
\end{tabular}
\end{footnotesize}
\caption{$\delta$ and $\Delta$ as the function of the number of steps}
\label{fig:diag}
\end{center}
\end{figure}

A typical result of the previously described algorithm can be seen in Figure \ref{fig:szor} and \ref{fig:diag}, the programming was made with Mathematica \cite{math8}, the CPU time of implementation is a few minutes. Besides the $DACM$, we keep track of the following quantities during the optimization:
$$\sigma=\sum_{i} \{ \textrm{2nd largest eigenvalue of } E_i\}$$
$$\delta=\sum_{i} (\Tr(E_i E_i)-\langle \Tr(E_i E_i)\rangle)^2$$
$$\Delta=\sum_{i\ne j} (\Tr(E_i E_j)-\langle \Tr(E_i E_j)\rangle)^2$$

We can conclude that
\begin{itemize}
       \item The DACM converges to the same value, independently from the initial state and from the particular realization of the annealing process, hence we are close to the optimum (left subfigure of Fig. \ref{fig:szor})
       \item $\sigma$ converges to 0, so the optimal POVM contains rank-one projections (right subfigure of Fig. \ref{fig:szor})
       \item $\delta$ converges to 0, so for the optimal POVM, $\Tr(E_i E_i)$ is a constant (left subfigure of Fig. \ref{fig:diag}).
       \item $\Delta$ converges to 0, so for the optimal POVM, $\Tr(E_i E_j) ~(i\ne j)$ is a constant (right subfigure of Fig. \ref{fig:diag}). This convergence is the slowest, and we can see from the jumps when the algorithm pushes the process to a different path with different angles. 
\end{itemize}

Thus, all the conditions in (\ref{SIC}) are fulfilled, but here we have different constants than in the SIC-POVM case. We can also obtain that the optimal POVM is quasi-orthogonal to the diagonal matrices.

\subsection{Obtaining an analytic result}\label{anal}

From the previous section we know that the optimal POVM contains rank one projectors, so we can use this property to achieve faster convergence.
Let us parametrize the POVM in the following way:
$$E_i=c |h_i\>\<h_i|,$$
where $c=3/7$, because $\sum_{i=1}^7 E_i=I$. So this way instead of parameterizing positive 3x3 matrices, we should parametrize 3 dimensional complex vectors:
$$|h_i\>=(z^{(1)}_i,z^{(2)}_i,z^{(3)}_i),$$ 
where  $\||h_i\>\|_2=1$.

On the other hand, we know from the previous section that $E_i$ is complementary to the diagonal subalgebra, i.e., it will have the same numbers in the diagonal, so we know that
\begin{equation}\label{cond_proj}
|z^{(1)}_i|^2=|z^{(2)}_i|^2=|z^{(3)}_i|^2=1/3, \quad (i=1, 2 \dots, 7). 
\end{equation}

Since $\Delta$ can be calculated easily directly from vectors,  $\<E_i,E_j\>=|\<h_i,h_j\>|^2$, 
we can minimize $\Delta$ numerically, using condition (\ref{cond_proj}). 
The convergence to zero is very fast, so we can conclude that there exists a POVM which is highly 
symmetric (satisfies the conditions in (\ref{SIC})) and minimizes (\ref{aim}).

\begin{pelda}\label{ex0}
By fixing some elements of the $|h_i\>$-s, we can get the following analytic result for the conditional SIC-POVM:

$$
E_1=\frac17 \left[
\matrix{
1&1&1\cr
1&1&1\cr
1&1&1}\right]
,
E_2=\frac17 \left[
\matrix{
1&\ep^6&\ep^2\cr
\ep&1&\ep^3\cr
\ep^5&\ep^4&1}\right]
,
E_3=\frac17 \left[
\matrix{
1&\ep^2&\ep^3\cr
\ep^5&1&\ep\cr
\ep^4&\ep^6&1}\right]
,$$
$$
E_4=\frac17 \left[
\matrix{
1&\ep^4&\ep^6\cr
\ep^3&1&\ep^2\cr
\ep&\ep^5&1}\right]
,
E_5=\overline{E_2}
,~
E_6=\overline{E_3}
,~
E_7=\overline{E_4}, 
$$
where $\ep=\exp\left(\frac{2\pi i}{7}\right)$.
\end{pelda}

%We can easily check that this POVM fulfills the following conditions: $\<E_i,E_i\>=9/49$ and $\<E_i,E_j\>= 2/49$. 
%So using the notations of (\ref{SIC}), we get constants $k=7$, $\lambda=7/3$, $\mu=2/9$. 

\subsection{Other examples}

We can use the algorithm described in Section \ref{alg} for other examples of state estimation problems, too.

It is clear that the same method can be used in 4 dimensions, the main difference is that there will be more grid points, therefore the clustering part will be longer. The second part of the algorithm, i.e., the search of the optimal POVM mainly depends on the number of unknown parameters, so if we do not have many unknown parameters the problem is not much more difficult than in the 3-dimensional case.

In the 4-dimensional case we can use the orthonormal basis
$$
\{\frac12~  \sigma_i\otimes \sigma_j\},\quad  i,j\in \{0,1,2,3\}
$$
for parametrization, where $\sigma_i~ (i\in \{0,1,2,3\})$ are the 2-dimensional Pauli-matrices.

\begin{pelda}\label{ex1}
If $\rho \in M_4(\bbbc)$ and we do not know the parameters related to \{$I\otimes \sigma_3$, $\sigma_3\otimes I$, $\sigma_3\otimes \sigma_3$\}, i.e., we want to estimate the diagonal elements of the density matrix, then the optimal POVM is $E_i=\left([A]_{j,k}=\delta_{i,j} \cdot \delta_{i,k}\right), i\in \{1,2,3,4\}$, i.e., the diagonal matrix units (i.e. rank-one projections).
\end{pelda}

\begin{pelda}\label{ex2}
If $\rho \in M_4(\bbbc)$ and we do not know the parameters related to \{$\sigma_1\otimes I$, $\sigma_2\otimes I$, $\sigma_3\otimes I$\}, i.e., we want to estimate $M_2\otimes I$, then the optimal POVM is $E_i=F_i \otimes I, i\in \{1,2,3,4\}$, where $F_i$-s are the elements of the 2-dimensional SIC-POVM. In this case $E_i=1/2 P_i$, where $P_i$ is a projection of rank two.
\end{pelda}

%So in this case we will have similar results than in the Section \ref{results}: the optimal measurement is symmetric ($\Tr E_i E_j=$constant) and $E_i$ is complementary to the known part. The only difference, that here the POVM elements are a multiples of a projection of rank two.

\begin{pelda}\label{ex3}
If $\rho \in M_4(\bbbc)$ and we do not know the parameters related to \{$\sigma_1\otimes I$, $\sigma_2\otimes I$, $\sigma_3\otimes I$, $I \otimes \sigma_1$, $I \otimes \sigma_2$, $I \otimes \sigma_3$\}, then the result of the algorithm indicates that the optimal POVM has the following properties: $E_1, E_2, E_3$ are in $M_2 \otimes I$, while $E_4, E_5, E_6$ are in $I \otimes M_2$ and all of them have the eigenvalues: $(\frac27,\frac27,0,0)$, so they are multiples of rank-2 projections. But $E_7$ has eigenvalues $(\frac27,\frac17,\frac17,0)$, so the optimal POVM does not contain only multiples of projections, and although one can observe some kind of symmetry,  $\Tr E_i E_j$ is not a constant.
\end{pelda}

\section{Conclusion and discussion}

The results from Theorem \ref{T1} and Example \ref{ex0} - \ref{ex2} show us that if some parameters of the quantum states are known, then the optimal POVM for state estimation has some appealing properties:
\begin{enumerate}
\item $E_i=c P_i$, where  $P_i$ is projection, $i\in \{1,2,\dots N+1\}$,
\item $\Tr E_i E_j=d, \quad i,j\in\{1,2,\dots N+1\}\quad$ and $\quad i\ne  j$,
\item $\Tr \sigma_j E_i=0, \quad  i\in\{1,2,\dots N+1\},\quad   j\in\{N+1,N+2, \dots, n^2-1\}$,
\end{enumerate}
using the abbreviations from Section \ref{optprob}. This means that the elements of the POVM are constant multiples of some projectors, symmetrical and complementary to the known part of the state. 

We will call the POVMs fulfilling conditions (i)-(iii) conditional SIC-POVMs, because they are symmetrical and informationally complete  with the condition of knowing some parameters of the state (hence we are not interested in distinguishing in those directions).
The conditional SIC-POVM is always defined with respect to the known parameters, since they define the characteristics of the optimal measurements through (iii). For instance, in Example \ref{ex1} and \ref{ex2} there are 4-dimensional states and the same number of unknown parameters, so one can expect the same result, however the optimal POVMs contain projections with different ranks (rank-1 and rank-2 respectively).

This is a generalization of SIC-POVMs, since we can get them as a special case using $N=n^2-1$ (that is, all parameters are unknown) and rank-one projections. Note that using this information, constants $c$ in (i) and $d$ in (ii) can be determined, while condition (iii) will result in an empty set. 

The existence of such measurements is a difficult question even in the unconditional case; the conjecture is that for every dimension there exists a SIC-POVM. The proposed algorithm shows us (Example \ref{ex3}) that for the conditional version the existence is not always true, however our results suggest that if the conditional SIC-POVM exists, then it is optimal.

We proposed an algorithm (Section \ref{alg}) that can be a useful tool to gain a better understanding of the optimal POVMs. It can be applied to higher dimensions problems, too, the only problem is the increased number of parameters, which will slow down the convergence, making unable to handle large systems. If we are only interested in the existence of a rank-one conditional SIC-POVM, we can use the algorithm in Section \ref{anal}, which converges much faster. 

Finally, we want to note that Example \ref{ex0} shows us, that the class of conditional SIC-POVMs is certainly not trivial, and the existence of conditional SIC-POVMs can be a fundamental question in different quantum tomography problems, therefore further investigations are suggested.

\end{document}